\documentclass[aps,pra,twocolumn,superscriptaddress]{revtex4-2}

\usepackage{float}

\usepackage{amsmath}
\usepackage{amssymb}
\usepackage{amsmath}
\usepackage{amsthm}
\usepackage{braket}
\usepackage{dsfont}
\usepackage{graphics}
\usepackage{graphicx}
\usepackage[hidelinks]{hyperref}
\usepackage{slashed}
\usepackage[titletoc]{appendix}
\usepackage{xcolor}
\usepackage{todonotes}
\definecolor{LightRed}{RGB}{255,235,235}
\definecolor{LightGreen}{RGB}{235,255,235}
\definecolor{LightBlue}{RGB}{235,235,255}
\definecolor{DarkRed}{RGB}{80,0,0}
\definecolor{DarkGreen}{RGB}{0,80,0}
\definecolor{DarkBlue}{RGB}{0,0,80}

\usepackage{comment}

\renewcommand{\vec}[1]{\boldsymbol{#1}}

\begin{document}
\newtheorem{my_definition}{Definition}

\title{Parameter space investigation for spin-dependent electron diffraction in the Kapitza-Dirac effect}
\author{Yang Wang}
\affiliation{Shanghai Normal University, Shanghai 200234, China}%
\affiliation{East China Normal University, Shanghai 200241, China}%
\author{Sven Ahrens}
\email{ahrens@shnu.edu.cn}
\affiliation{Shanghai Normal University, Shanghai 200234, China}%

\begin{abstract}
We demonstrate that spin-dependent electron diffraction is possible for a smooth range of transverse electron momenta in a two-photon Bragg scattering scenario of the Kapitza-Dirac effect. Our analysis is rendered possible by introducing a generalized specification for quantifying spin-dependent diffraction, yielding an optimization problem which is solved by making use of a Newton gradient iteration scheme. With this procedure, we investigate the spin-dependent effect for different transverse electron momenta and different laser polarizations of the standing light wave Kapitza-Dirac scattering. The possibility for using arbitrarily low transverse electron momenta, when setting up a spin-dependent Kapitza-Dirac experiment allows longer interaction times of the electron with the laser and therefore enables less constraining parameters for an implementation of the effect.
\end{abstract}
\maketitle

\section{Introduction}

In 1933, Kapitza and Dirac have predicted the reflection of electrons at standing waves of light \cite{kapitza_dirac_1933_proposal}, known as the Kapitza-Dirac effect, nowadays. The effect has been demonstrated experimentally for atoms \cite{gould_1986_atoms_diffraction_regime,martin_1988_atoms_bragg_regime} in 1986, \cite{gould_1986_atoms_diffraction_regime,martin_1988_atoms_bragg_regime} and then also for electrons \cite{Bucksbaum_1988_electron_diffraction_regime}, where another precise experiment was conducted in 2001 and 2002, which was confirming the Kapitza-Dirac effect with multiple \cite{Freimund_Batelaan_2001_KDE_first} and single \cite{Freimund_Batelaan_2002_KDE_detection_PRL} diffraction orders for electrons. The experimental realization of the Kapitza-Dirac effect raised the question about spin effects \cite{Batelaan_2003_MSGE,rosenstein_2004_first_KDE_spin_calculation}, which were suggested theoretically \cite{ahrens_bauke_2012_spin-kde,ahrens_bauke_2013_relativistic_KDE,bauke_ahrens_2014_spin_precession_1,bauke_ahrens_2014_spin_precession_2,erhard_bauke_2015_spin}. Indeed, a recent implementation of the Kapitza-Dirac effect on the basis of transmission electron microscopy is showing an expected dip of the diffracted beam at the location where spin effects are expected, provided that sufficient experimental accuracy was available \cite{Axelrod_2020_Kapitza_Dirac_cancellation_observation}. Follow-up theoretical investigations focused on bi-chromatic laser configurations to show spin-effects in the Kapitza-Dirac effect \cite{McGregor_Batelaan_2015_two_color_spin,dellweg_awwad_mueller_2016_spin-dynamics_bichromatic_laser_fields,dellweg_mueller_2016_interferometric_spin-polarizer,dellweg_mueller_2016_interferometric_spin-polarizer,dellweg_mueller_extended_KDE_calculations,ebadati_2018_four_photon_KDE,ebadati_2019_n_photon_KDE} and refined descriptions suggested coherent electron spin polarization and spin inference by the interaction with light only \cite{Karlovets_2011_radiative_polarization,Del_Seipt_Blackburn_Kirk_2017_electron_spin_polarization,Wen_Tamburini_Keitel_2019_polarized_kilo_ampere_beams,van_Kruining_Mackenroth_2019_magnetic_field_polarization,Chen_Keitel_2019_polarized_positrons,Li_Chen_2020_polarized_positron_electron,Xue_Chen_Hatsagortsyan_Keitel_Li_2022_GeV_polarization,Gong_Hatsagortsyan_Keitel_2023_plasma_instability_polarization}.

Among the predictions of a spin-dependent Kapitza-Dirac effect, several scenarios consider three or more photon interactions \cite{ahrens_bauke_2012_spin-kde,ahrens_bauke_2013_relativistic_KDE,McGregor_Batelaan_2015_two_color_spin,dellweg_awwad_mueller_2016_spin-dynamics_bichromatic_laser_fields,dellweg_mueller_extended_KDE_calculations,ebadati_2018_four_photon_KDE,ebadati_2019_n_photon_KDE} or a quantum state evolution up to at least fractions of the transition's Rabi cycle are required \cite{dellweg_mueller_2016_interferometric_spin-polarizer,ahrens_2017_spin_filter}. A recently discussed, two-photon interaction in a Bragg scattering setup opens the perspective for an experimental implementation of the effect at hard X-ray standing light waves \cite{ahrens_2020_two_photon_bragg_scattering}, but the necessary electron momentum of $1mc$ puts limits on the interaction time of the laser with the electron. With low electron velocities one could achieve higher spin-dependent diffraction probabilities, which would ease the necessary demand for the peak laser intensity of the experiment.

Nevertheless, the parameters discussed in reference \cite{ahrens_2020_two_photon_bragg_scattering} are not the only possible set of parameters for a spin-dependent two-photon interaction in the Kapitza-Dirac effect. From the Taylor expansion of the spin- and polarization dependent scattering matrix in \cite{ahrens_2020_two_photon_bragg_scattering} one can see that similar spin-dependent two-photon diffraction setups will be possible, when smoothly varying the setup parameters. Consequently, we investigate different parameter ranges, for observing the spin-dependent electron diffraction effect. We do this by establishing a general formulation for the characterization of spin-dependent electron diffraction, where we implement an iterative algorithm for the optimization of spin-dependent diffraction by using the Newton method in two dimensions. Being equipped with this tool, we are able to demonstrate spin-dependent electron diffraction, even for vanishing zero transverse electron momenta.

Our article is structured as follows. In section \ref{sec:theory_description}, we present our theory. We begin with introducing the laser field, the electron quantum state and the related Compton scattering formula for the two-photon Kapitza-Dirac effect in the Bragg regime in section \ref{sec:compton_scattering}. We then discuss and specify the meaning of \emph{spin-dependent} diffraction in the context of the Kapitza-Dirac effect in section \ref{sec:spin_dependence_discussion} and introduce a quantity which we call 'contrast', for quantifying this spin-dependence in section \ref{sec:contrast}. We then demonstrate the functionality of an iterative algorithm for determining the optimized spin parameters for the contrast in section \ref{sec:parameter_study}. We do this first in the context of a known literature example (section \ref{sec:method_demonstration}) and then show that the optimization algorithm can be used for smoothly lowering the transverse electron momentum to zero (section \ref{sec:smooth_momentum_change}). In section \ref{sec:discussion}, we verify the algorithmically determined results by comparing the results with analytic solutions on the basis of a Taylor expansion of the Compton tensor. We finally present an outlook for the use of the our method in section \ref{sec:outlook} and provide a documentation of the algorithmic implementation for the contrast optimization procedure in Appendix \ref{sec:iterative_algorithm}. We also provide a refined description about the spin configuration of the electron when undergoing scattering in Appendix \ref{sec:spin_expectation_value}.

\section{Theoretical Description\label{sec:theory_description}}
\subsection{Electron diffraction and Compton scattering\label{sec:compton_scattering}}

In our investigation, we consider the electron diffraction at a standing wave laser beam, which is propagating along the $x$-axis
\begin{subequations}%
\begin{align}%
 \vec A_{r}&=\vec A_{0,r}e^{i(\omega t-k_L x)}+\vec A_{0,r}^{\ast}e^{-i(\omega t-k_L x)}\label{laser_r}\\
 \vec A_{l}&=\vec A_{0,l}e^{i(\omega t+k_L x)}+\vec A_{0,l}^{\ast}e^{-i(\omega t+k_L x)}\,,\label{laser_l}
\end{align}\label{eq:laser_beam}%
\end{subequations}%
where $\vec A_{r}$ ($\vec A_{l}$) is the polarization of the beam travelling in positive (negative) direction along $x$-axis, with the amplitude $\vec A_{0,r}$ ($\vec A_{0,l}$), respectively. Denoted are also the laser wave number $k_L$, laser frequency $\omega=k_L$ and time $t$, where we are setting $\hbar=c=1$ in this article, with the exception of exemplifying the transverse electron momentum in terms of keV/$c$ and the laser photon energy in terms of eV in section \ref{sec:method_demonstration}.% The symbol $*$ denotes complex conjugation.

For the initial electron quantum state in the two-photon Kapitza-Dirac effect we assume the wave function
\begin{subequations}%
\begin{equation}%
\Psi_i(\vec x,t_0)=\sum_{s}c_{i}^{s}(t_0)u_{\vec p_i}^{+,s}e^{i\vec x \cdot \vec p_i}
\label{eq:initial_wave_function}
\end{equation}%
at initial time $t_0$ and for the electron after the interaction with the laser we similarly write the final quantum state
\begin{equation}%
\Psi_f(\vec x,t)=\sum_{s}c_{f}^{s}(t)u_{\vec p_f}^{+,s}e^{i\vec x \cdot \vec p_f}\,.
\label{eq:final_wave_function}
\end{equation}\label{eq:wave_functions}%
\end{subequations}%
Initial and final electron momenta
\begin{subequations}%
\begin{align}%
 \vec p_i = - k_L \vec e_1 + p_2 \vec e_2 + p_3 \vec e_3\\
 \vec p_f = \phantom{-} k_L \vec e_1 + p_2 \vec e_2 + p_3 \vec e_3
\end{align}%
\end{subequations}%
are chosen such that energy and momentum conservation are fulfilled \cite{ahrens_2012_phdthesis_KDE,ahrens_bauke_2012_spin-kde,ahrens_bauke_2013_relativistic_KDE}, for the case of absorption of one photon from the right propagating laser beam and the induced emission of another photon into the left propagating laser beam, from the external field \eqref{eq:laser_beam}. In Eqs. \eqref{eq:wave_functions} we also denote the bi-spinors
\begin{subequations}%
\begin{align}%
u_{\vec p}^{+,s}&=\sqrt{\frac{E_{\vec p}+m}{2m}}
\begin{pmatrix}
\chi^s\\
\frac{\vec \sigma \cdot \vec p}{E_{\vec p}+m}\chi^s
\end{pmatrix}
\label{eq:bi-spinors_up}\\
u_{\vec p}^{-,s}&=\sqrt{\frac{E_{\vec p}+m}{2m}}
\begin{pmatrix}
\frac{\vec \sigma \cdot \vec p}{E_{\vec p}+m}\chi^s\\
\chi^s
\end{pmatrix}\,,
\label{eq:bi-spinors_down}
\end{align}\label{eq:bi-spinors}%
\end{subequations}%
with the vector $\vec \sigma$ of Pauli matrices
\begin{equation}
\sigma_{1}=
\begin{pmatrix}
0&1\\
1&0
\end{pmatrix}\,,\ 
\sigma_{2}=
\begin{pmatrix}
0&-i\\
i&0
\end{pmatrix}\,,\ 
\sigma_{3}=
\begin{pmatrix}
1&0\\
0&-1
\end{pmatrix}
\label{eq:pauli_matrices}
\end{equation}
and relativistic energy momentum relation
\begin{align}
E_{\vec{p}}=&\sqrt{m^{2}+p^{2}}\,,\label{eq:relativistic_energy_momentum_relation}
\end{align}
where $m$ is the electron rest mass. %The central dot between two three vectors denotes the inner product in Euclidean space.
Note, that we are referring to the spatial direction of the $\vec e_1$, $\vec e_2$, $\vec e_3$ unit vectors, when mentioning the $x$-, $y$- and $z$-axis, in this article.

The spin-dependent quantum state propagation from the initial to the final electron quantum state in the event of scattering at the laser beam can be written as
\begin{equation}
c_{f}^{s'}(t)=\sum_{s}U^{s',s}(t,t_0)c_{i}^{s}(t_0)\,.
\label{eq:propagation_equation}
\end{equation}
For the mentioned case of absorption and emission of a single photon (two photon interaction) in a standing wave laser beam, we have shown that the diffraction probability is proportional to the Compton scattering formula \cite{ahrens_2020_two_photon_bragg_scattering}
\begin{multline}
 U^{s',s}(t,t_0) \sim M^{s s'}= (A_{0,l}^*)_\mu (A_{0,r})_\nu \Tilde{M}^{s's;\mu\nu}\,,
\label{Compton_formula}
\end{multline}
with the Compton tensor
\begin{multline}
 \Tilde{M}^{s's;\mu\nu} \\
 =\bar u^{s}_{\vec p_f} \left(  \gamma^\mu \frac{ \slashed p_i + \slashed k + m }{2 p_i \cdot k} \gamma^\nu - \gamma^\nu \frac{ \slashed p_i - \slashed k' + m }{2 p_i \cdot k'} \gamma^\mu \right) u^{s'}_{\vec p_i}\,.
\label{Compton_tensor}
\end{multline}
Here we are denoting the four vectors $p_i = (E_{\vec{p_i}},\vec p_i)$, $k = (\omega,\vec k)$ and $k'= (\omega,-\vec k)$, the photon momentum $\vec k = (0,0,k_L)$ and Dirac adjoint $\bar u^{s}_{\vec p} = u^{s\dagger}_{\vec p} \gamma^0$. The expressions are based on Einstein's sum convention with metric $g=\textrm{diag}(1,-1,-1,-1)$ and Dirac gamma matrices
\begin{equation}
 \gamma^0=
\begin{pmatrix}
 \mathds{1} & 0\\
 0 &\mathds{1}
\end{pmatrix}
\,,\qquad
\gamma^i =
\begin{pmatrix}
0 & \sigma^i\\
-\sigma^i & 0
\end{pmatrix}\,,
\end{equation}
where $\mathds{1}$ is the $2\times 2$ identity. Further details about conventions in quantum field theory can be found, for example in references \cite{Peskin_Schroeder_1995_Quantum_Field_Theory,Landau_Lifshitz_1982_Quantum_Electrodynamics,Ryder_1986_Quantum_Field_Theory,Srednicki_2007_Quantum_Field_Theory,Halzen_Martin_1984_quarks_and_leptons,Weinberg_1995_quantum_theory_of_fields}.

We introduce the dimensionless parameters
\begin{equation}
 q_L = \frac{k_L}{m}\,,\qquad
 q_2 = \frac{p_2}{m}\,,\qquad
 q_3 = \frac{p_3}{m}\,,
\end{equation}
in place of the laser wave number $k_L$ and the transverse momenta $p_2$ and $p_3$ of the electron, for ease of notion, in the following. Note, that the word `transverse' is used with respect to the laser beam propagation direction ($x$-direction), in this article.

\subsection{Characterization of spin-dependent diffraction\label{sec:spin_dependence_discussion}}

We have introduced the spin propagation matrix $M$ as $S$-matrix component of Compton scattering in Eq. \eqref{Compton_formula}, which depends on a multidimensional space of physical parameters, ie. polarization and momenta of the laser beams and the electron. Therefore, the exact form of $M$ is difficult to predict and we thus assume the matrix $M$ to be a general complex $2\times 2$ matrix, $M \in \mathbb{C}^{2\times 2}$ with 8 independent degrees of freedom in the following discussion. After having established the concept of the `contrast' below, we illustrate our formalism again with specific matrix entries $M$ from Eq. \eqref{Compton_formula} in sections \ref{sec:parameter_study} and \ref{sec:discussion}.

We point out, that we base our investigation directly on the complex entries of the $S$-matrix, as in Eqs. \eqref{Compton_formula} and \eqref{Compton_tensor} in this work and it might be fruitful, to understand how our `plain matrix' treatment might generalize in spin parameterizations on the basis of the stokes vector \cite{Torgrimsson_2019_non_linear_compton_scattering,Torgrimsson_2020_nonlinear_qed_approximation,Torgrimsson_2021_loops_polarization} or the spin density matrix \cite{Seipt_2018_radiative_electron_polariztion} in further studies, for details see for example \cite{King_Ilderton_Seipt_Torgrimsson_Karbstein_Fedotov_2023_strong_field_qed_review}. In contrast to a plane-wave external field situation, for which a solution of the Dirac equation is available in terms of the Volkov solution \cite{King_Seipt_2020_lcfa}, we base the matrix input for our considerations in the context of a standing light wave on a situation where the quantum dynamics is linear in the external fields, such that the Compton scattering formula can be used to describe the dynamics \cite{ahrens_2020_two_photon_bragg_scattering}. Note, that our focus in this work is the influence of the initial electron polarization on the final diffraction pattern after laser-electron interaction. The classical electron motion can also be influenced by a spin dependence of the radiation reaction force \cite{Del_Seipt_Blackburn_Kirk_2017_electron_spin_polarization}. In this context, it is possible, to describe electron spin-polarization by including photon emissions along a semi-classical particle motion in a Monte Carlo implementation \cite{Keitel_Li_2019_single_shot_polarization}.

Even though we have already identified projection matrices as one possible characterization criterion for spin-dependent electron diffraction in the Kapitza-Dirac effect in a previous investigation \cite{ahrens_2017_spin_filter}, more general matrix forms for spin-dependent diffraction are possible \cite{ahrens_2020_two_photon_bragg_scattering}.

At this point we want to emphasize that we associate the term \emph{spin-dependent} diffraction as a diffraction pattern which depends on the initial spin state of the electron, in context of the Kapitza-Dirac scattering. The large number of matrix degrees of freedom for the electron spin-propagation matrix turn the question ``How can one define a general and unique characterization for spin-dependent spin propagation matrices?'' into a sophisticated problem.

As a first step, one may be tempted to establish a general definition for the term `spin-dependent spin propagation matrix'. For such a definition it stands to reason to require the following two conditions:\\

\noindent Spin-dependent spin propagation matrices should be
\begin{enumerate}
 \item non-zero and
 \item have a non-zero kernel dimension.
\end{enumerate}
This definition would guarantee diffraction, as the matrix is required to be non-vanishing (first condition). The diffraction will also be spin-dependent, because we have defined the guaranteed existence of an electron spin polarization which will be mapped on the zero vector, which is the kernel of the matrix (second condition).

The next question, which arises, is how one would handle this definition in practical terms, in a numeric implementation. The accuracy of the numeric representation of numbers turns the second condition into non-trivial problem, and this problem is related to another question, regarding spin-dependent diffraction: What about diffraction which is only partially spin-dependent? To specify the term `partially spin-dependent', we assume two orthogonal electron spin polarizations $\psi^A \in \mathbb{C}^{2}$ and $\psi^B \in \mathbb{C}^{2}$. Specifically, in this article we choose the commonly known states
\begin{equation}\label{bloch_states}
\psi^A =
\begin{pmatrix}
\cos\left(\frac{\alpha}{2}\right)\\
\sin\left(\frac{\alpha}{2}\right) e^{i \varphi}
\end{pmatrix}\,, \qquad
\psi^B =
\begin{pmatrix}
\sin\left(\frac{\alpha}{2}\right)e^{-i \varphi}\\
- \cos\left(\frac{\alpha}{2}\right)
\end{pmatrix}\,,
\end{equation}
whose expectation value with respect to the vector of spin matrices $\vec \sigma$ points at direction $\vec n = (\sin \alpha \cos \varphi, \sin \alpha \sin \varphi,\cos \alpha)$ and $- \vec n$ on the Bloch unit sphere, respectively.

\subsection{Quantifying spin-dependent diffraction\label{sec:contrast}}

As a next step in quantifying the term `partially spin-dependent', we define a quantity which we call `contrast' as the fraction
\begin{equation}
  \mathcal{C}'(M) = \frac{|M \psi^A|^2}{|M \psi^B|^2}\label{contrast}
\end{equation}
evaluated for a given matrix $M$ at the value pair $(\alpha,\varphi)$, such that $\mathcal{C}'(M)$ is minimal. In other words, the contrast $\mathcal{C}(M)$ of the matrix $M$ is the minimum value of $\mathcal{C}'(M)$, where the minimum is to be determined with respect to the variables $\alpha$ and $\varphi$ of the spinors in Eq. \eqref{bloch_states}. In the context of this definition the stated criterion for a spin-dependent spin propagation matrix as a matrix which is non-vanishing and has a non-zero kernel dimension corresponds to a spin propagation matrix with zero contrast $\mathcal{C}(M)=0$. On the other hand, the maximally possible value for the contrast emerges for a situation, in which $|M \psi^A|^2=|M \psi^B|^2$, since the requirement for a minimum of $\mathcal{C}'(M)$ implies that a value pair $(\alpha,\varphi)$ is chosen for which $|M \psi^A|^2$ is smaller or equal $|M \psi^B|^2$. Therefore, the maximally possible value for the contrast is one and we have $0 \le \mathcal{C}(M) \le 1$, for all $M \in \mathbb{C}^{2\times 2}$. Within the frame work of this description, partial spin-dependent diffraction corresponds to a contrast which is larger than zero and which is smaller than one.

A last question in the discussion about spin-dependence arises about the determination of the values $\alpha$ and $\varphi$, at which $\mathcal{C}'(M)$ is minimal. An exact, constructive method for the explicit determination for the value pair $(\alpha,\varphi)$ for the minimum might exist. Nevertheless, in this work, we are pursuing a pragmatic approach, by implementing a Newton method in the two-dimensional space of the variables $\alpha$ and $\varphi$ for finding a local minimum of $\mathcal{C}'(M)$. The implementation details of the Newton procedure in our work are summarized in appendix \ref{sec:iterative_algorithm} and an executable example for running the optimization algorithm is provided in the supplemental material.

\section{Parameter study of spin-dependent diffraction\label{sec:parameter_study}}

\subsection{Method demonstration\label{sec:method_demonstration}}

\begin{figure}
  \includegraphics[width=0.5\textwidth]{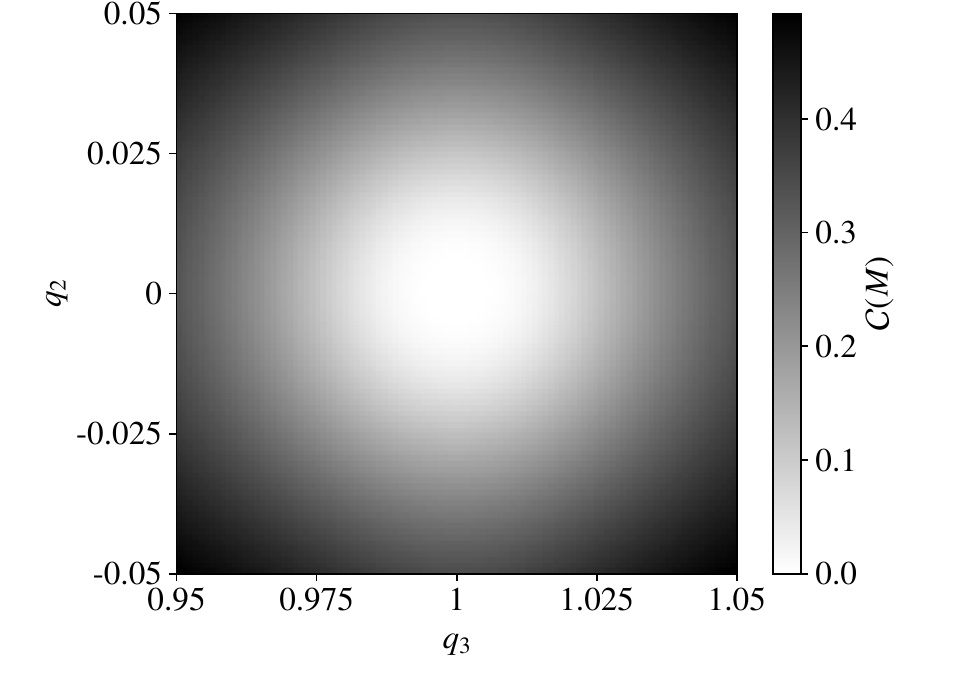}
  \caption{Example for a vanishing contrast. The contrast $C(M)$ is computed for the scenario described in section \ref{sec:method_demonstration}, and evaluated according to the procedure as discussed in section \ref{sec:contrast}, where the spin propagation matrix $M$ is based on the Compton scattering formula \eqref{Compton_formula}. The iterative method for the determination of $C(M)$ is described in appendix \ref{sec:iterative_algorithm}. We evaluate $M$ and $C(M)$ as a function of the transverse electron momenta $q_2$ and $q_3$, around the centering values $q_2 = 0$ and $q_3 = 1$, according to parameters similar to the parameters in reference \cite{ahrens_2020_two_photon_bragg_scattering}. Also, as in references \cite{ahrens_2020_two_photon_bragg_scattering,ahrens_2017_spin_non_conservation}, we choose the laser energy $q_L=0.02$ and the laser polarizations $\vec A_{0,r} = \vec e_3$ and $\vec A_{0,l} = (\vec e_2 + i \vec e_3)/\sqrt{2}$. \label{con_095_105}}
\end{figure}

\begin{figure}
  \includegraphics[width=0.5\textwidth]{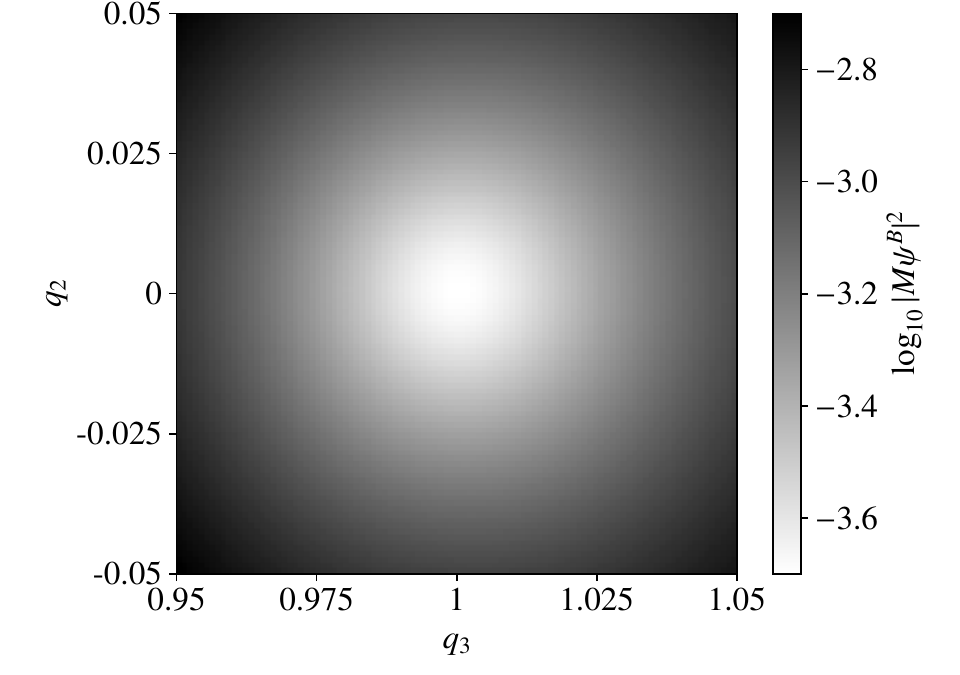}
\caption{Display of a non-vanishing diffraction probability for the transverse momentum parameters as in Fig. \ref{con_095_105}. In analogy to the procedure in Fig. \ref{con_095_105}, we compute the spin propagation matrix $M$ as function of the transverse momenta $q_2$ and $q_3$. Different to Fig. \ref{con_095_105} however, we display the maximally possible diffraction probability $|M\psi^B|^2$, in place of the contrast $C(M)$. For each value pair $(q_2,q_3)$ in the density plot, we use the iteratively optimized values $(\alpha,\varphi)$ from our algorithm for the determination of $C(M)$. We point out that the circumstance of a zero contrast in Fig. \ref{con_095_105} in combination with a non-vanishing diffraction probability in this figure implies that spin-dependent diffraction can emerge for suitable choices of $\alpha$ and $\varphi$, in the Kapitza-Dirac effect, in accordance with the conclusion drawn in references \cite{ahrens_2020_two_photon_bragg_scattering,ahrens_2017_spin_non_conservation}.
\label{M_095_105}}
\end{figure}

We now want to apply the above discussed formalism to a specific example. From previous investigations, a spin-dependent electron-laser interaction is predicted in a standing light wave, in which one of the counterpropagating laser beams is linearly polarized, while the other beam is circularly polarized, and additionally the electron has a transverse momentum of 511\,keV$/c$ along the polarization direction of the linearly polarized beam \cite{ahrens_2020_two_photon_bragg_scattering,ahrens_2017_spin_non_conservation}. We therefore set the polarization vectors of the laser beam \eqref{eq:laser_beam} to the values $\vec A_{0,r} = \vec e_3$, $\vec A_{0,l} = (\vec e_2 + i \vec e_3)/\sqrt{2}$, compute the matrix $M$ according to the Compton scattering formula \eqref{Compton_formula} and determine the contrast $C(M)$ as a function of the two transverse electron momenta around the values $q_2 = 0$ and $q_3 = 1$ in Fig. \ref{con_095_105}. Note, that the value $q_3 = 1$ corresponds to the above mentioned momentum 511\,keV$/c$. Also, we set the wave number of the laser beam to the value $q_L = 0.02$, in accordance to the corresponding value in related investigations \cite{ahrens_2017_spin_non_conservation,ahrens_2020_two_photon_bragg_scattering,ahrens_guan_2022_beam_focus_longitudinal}. The value $q_L = 0.02$ corresponds to X-ray photons with an energy of about 10220\,eV. The resulting contrast in Fig. \ref{con_095_105} is indeed approaching zero, for parameters at which spin-dependent diffraction is already predicted in Ref. \cite{ahrens_2020_two_photon_bragg_scattering}. Therefore, Fig. \ref{con_095_105} illustrates the non-zero kernel dimension for the occurrence of spin-dependent diffraction, as it is discussed above. The other mentioned criterion, which is a non-zero spin propagation matrix for observing spin-dependent diffraction, can be verified by another plot as shown in Fig. \ref{M_095_105}. Fig. \ref{M_095_105} contains the same parameters as in Fig. \ref{con_095_105}, but now the quantity $|M \psi^B|^2$ is shown, instead of $C(M)$. The non-vanishing probability $|M \psi^B|^2$ in Fig. \ref{M_095_105} implies that the initial spin polarization $\psi^B$ still results in a final diffraction probability. This can only happen, if the spin propagation matrix is non-zero. In summary, the data in Figs. \ref{con_095_105} and \ref{M_095_105} agrees with the expectation that electrons with initial polarization $\psi^A$ are not (or only less) diffracted, while electrons with polarization $\psi^B$ are undergoing diffraction. We therefore have successfully applied our theory framework to a situation, in which the final diffraction pattern of the Kapitza-Dirac effect depends on the choice of the initial electron spin polarization (either $\psi^A$ or $\psi^B$). Note, that we provide a comprehensive description of the electron spin state before and after Kapitza-Dirac scattering for Figs. \ref{con_095_105} and \ref{M_095_105} in Appendix \ref{sec:spin_expectation_value}.

\subsection{Spin-dependent diffraction on smooth parameter change\label{sec:smooth_momentum_change}}

In the following, we are interested in finding parameters for the laser polarization and electron momentum, for which the electron undergoes spin-dependent diffraction in Kapitza-Dirac scattering with having a low transverse electron momentum. In this context, we mention that the polarization amplitudes $\vec A_{0,r}$ and $\vec A_{0,l}$ in Eq. \eqref{eq:laser_beam} are complex three component vectors, where the vacuum Maxwell equations imply that the component along the laser propagation axis vanishes, for a plane wave laser field. Therefore, for our beams propagating along the $x$-axis, the first component of $\vec A_{0,r}$ and $\vec A_{0,l}$ is zero, resulting in 8 remaining degrees of freedom (two complex numbers for each of the two polarization amplitudes) for the polarization of the standing wave laser beam.

The general exploration of the parameter space with respect to possible polarizations and electron momenta might be interesting. However, for the research question of this work, we specifically find that spin-dependent electron diffraction with low transverse electron momenta can be achieved by varying the ellipticity of one laser beam the two counter propagating laser beams, while keeping the other beam linearly polarized. We therefore choose the polarization
\begin{equation}
 \vec A_{0,l}=
 \begin{pmatrix}
     0\\
   \cos\theta\\
   i \sin\theta
 \end{pmatrix}
\,,\quad
 \vec A_{0,r}=
 \begin{pmatrix}
  0\\ 0\\ 1
 \end{pmatrix}\,.
\label{amplitude_r}
\end{equation}
The external field is compatible with the polarization used in Figs. \ref{con_095_105} and \ref{M_095_105} of the previous section, in case we set the ellipticity parameter to the value $\theta=\pi/4$ in Eq. \eqref{amplitude_r}.

\begin{figure}
 \includegraphics[width=0.5\textwidth]{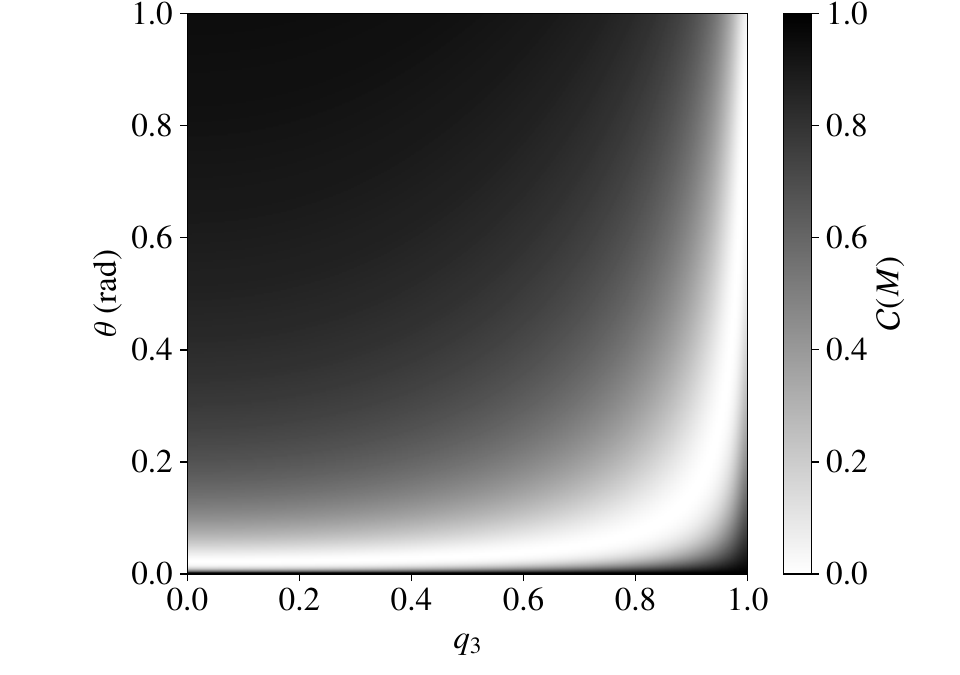}
 \caption{Contrast for a left propagating laser beam with elliptical polarization. Parameters and computation procedure are the same as in Fig. \ref{con_095_105}, except that we use the more general polarization \eqref{amplitude_r} with elliptical polarization angle $\theta$ and keep value for the transverse momentum $q_2$ fixed at value $0$. One can see that for every transverse momentum $q_3$ ($x$-axis) there is a polarization with angle $\theta$ ($y$-axis), for which the contrast reaches zero.
 \label{theta_p3}}
\end{figure}

In order to explore the occurrence of spin-dependent electron diffraction in dependence of the transverse electron momentum $q_3$ and the ellipticity parameter $\theta$ of the left propagating laser beam, we display the contrast $C(M)$ as a function of these two parameters in a density plot in Fig. \ref{theta_p3}. The parameter $q_2$, which was varied in Figs. \ref{con_095_105} and \ref{M_095_105}, is set to the constant value $q_2 = 0$. We see in Fig. \ref{theta_p3} that for each value $q_3$ there is an angle $\theta$, for which the contrast is zero, at the white regions in the density plot. This means, that for every transverse momentum $q_3$ in the displayed range $q_3 \in [0,1]$ one expects to find parameters for spin-dependent diffraction, provided the spin-independent diffraction probability at those parameters is non-vanishing.

\begin{figure}
 \includegraphics[width=0.487\textwidth]{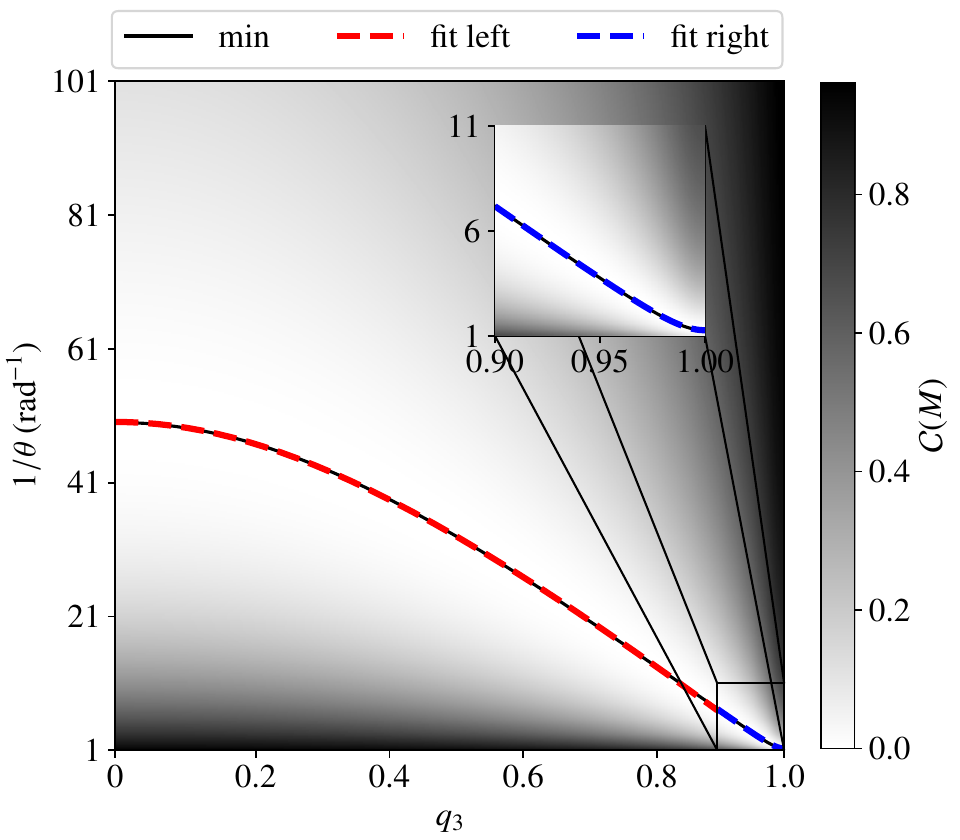}
 \caption{\label{1/theta_p3}The contrast as a function of the inverse ellipticity parameter $1/\theta$ and $q_3$. In this figure, we have inverted $\theta$ at the $y$-axis, as compared to Fig. \ref{theta_p3}, such that the minimum value of the contrast can be located easier. The location of the minimum is plotted in the figure as solid black line. We can fit this minimum location by the functions \eqref{left_fit} for the range $q_3 \in [0,0.9]$ (red dashed line) and \eqref{right_fit} for the range $q_3 \in [0.9,1]$ (blue dashed line). The inset is magnifying the fit of the blue dashed line. The value of $1/\theta$  at the left ($q_3=0$) and right ($q_3=1$) bounds of the figure match analytic predictions for a zero contrast, as shown in sections \ref{sec:consistency_considerations} and \ref{sec:low_electron_momentum}.}
\end{figure}

The choice of presentation for the contrast in Fig. \ref{theta_p3} appears unsuitable, as the values for a zero contrast are located in a region very close to $\theta=0$, but not exactly at $\theta=0$. For a better observation of the low contrast regions at $q_3=0$, we therefore display $\mathcal{C}(M)$ again as a function of $\theta$ and $q_3$ in Fig. \ref{1/theta_p3}, where we now use the inverse function $1/\theta$ instead of $\theta$ as $y$-axis for the density plot. The most interesting region in Fig. \ref{1/theta_p3} is the location where the contrast $\mathcal{C}(M)$ is has its lowest value. We denote this lowest contrast location as the function $\theta(q_3)$. The position of this numerically determined minimum $\theta(q_3)$ is marked by the black solid line `min' in Fig. \ref{1/theta_p3}. We find that the function can be approximated by the fitting function
\begin{subequations}\label{eq:fit_functions}%
\begin{equation}%
 \frac{1}{\theta}(q_3)=a_1+a_2 \sqrt{q_3^2+a_3}\label{left_fit}
\end{equation}%
on the interval $q_3 \in [0,0.9]$ and
\begin{equation}%
 \frac{1}{\theta}(q_3)=b_1+b_2 \sqrt{(q_3-1)^2+b_3}\label{right_fit}
\end{equation}%
\end{subequations}%
on the interval $q_3 \in [0.9,1]$. The fitting parameters for the functions are determined as
\begin{subequations}\label{eq:fit_parameters}%
\begin{align}%
a_1&=\phantom{-}9.671\times 10^{1} & 
b_1&=2.771 \times 10^{-2}\\
a_2&=-8.510 \times 10^{1} & 
b_2&=7.041 \times 10^{1}\\
a_3&=\phantom{-}2.996\times 10^{-1} &
b_3&=3.137 \times 10^{-4}\,.
\end{align}%
\end{subequations}%
We display the fitting functions \eqref{eq:fit_functions} with parameters \eqref{eq:fit_parameters} as dashed red and dashed blue lines in Fig. \ref{1/theta_p3} for demonstrating that they approximate the minimum function $\theta(q_3)$ well.

% exact fit parameters
% \begin{flalign}
%   a_1&=\phantom{-}9.671300918\times 10^{1} & b_1&=2.77095187 \times 10^{-2}\nonumber\\
%   a_2&=-8.519649553 \times 10^{1} & b_2&=7.04119167 \times 10^{1}\label{eq:fit_parameters}\\
%   a_3&=\phantom{-}2.9957552\times 10^{-1} & b_3&=3.13709980 \times 10^{-4}\,.\nonumber
% \end{flalign}
% 
% old fit parameters
% \begin{flalign}
%   a_1&=\phantom{-}9.671301044\times 10^{1} & b_1&=2.77095253\times 10^{-2}\nonumber\\
%   a_2&=-8.519649637\times 10^{1} & b_2&=7.04119167\times 10^{1}\label{eq:fit_parameters}\\
%   a_3&=\phantom{-}2.9957553\times 10^{-1} & b_3&=3.13709977\times 10^{-4}\,.\nonumber
% \end{flalign}

\begin{figure}
 \includegraphics[width=0.48\textwidth]{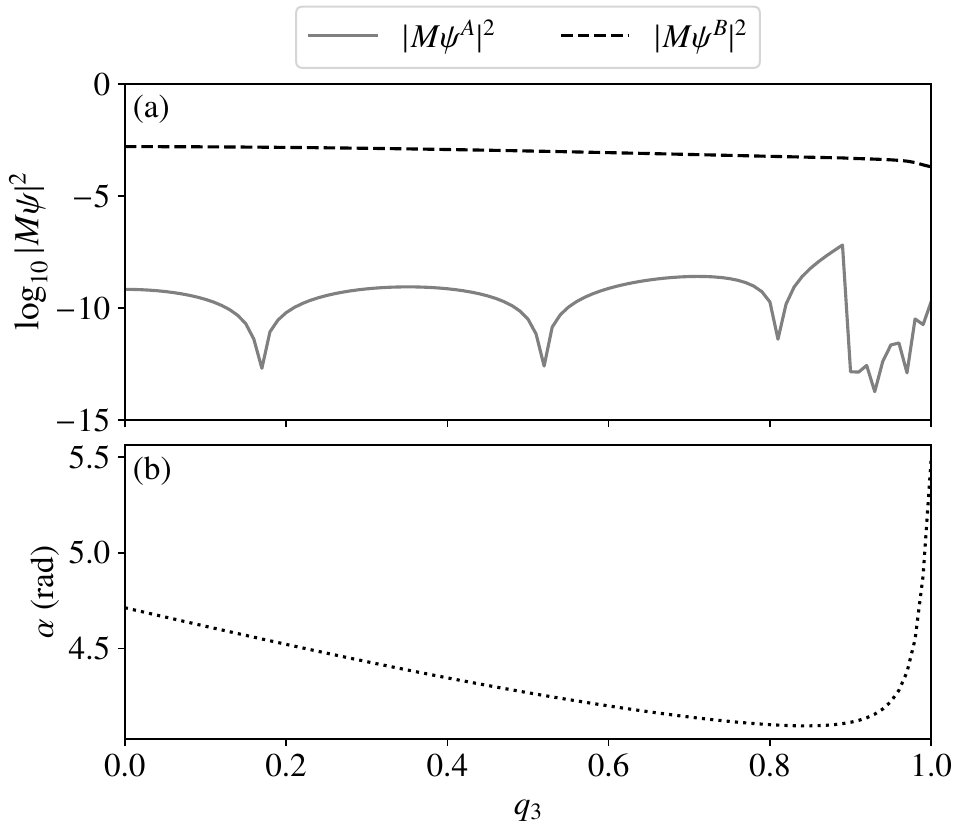}
 \caption{Explicit demonstration of a vanishing contrast along the dashed line in Fig. \ref{1/theta_p3}(a). We determine the parameter pair $(\alpha,\varphi)$ with our iterative contrast determination algorithm for the combination of $q_3$ and $1/\theta(q_3)$ along the minimum fit \eqref{eq:fit_functions} of the contrast distribution in Fig. \ref{1/theta_p3}. The probability $|M \psi^A|^2$ is several orders of magnitude smaller than $|M \psi^B|^2$ along the minimum location in Fig. \ref{1/theta_p3}, which implies a value of the contrast in Eq. \eqref{contrast} significantly smaller than 1. In combination with a non-vanishing amplitude for $|M \psi^B|^2$ this represents spin-dependent diffraction. Note, that tiny discrepancies between the fitting functions Eq. \eqref{eq:fit_functions} and the exact location of the contrast minimum in the $(1/\theta$,$p_3$) plane of Fig. \ref{1/theta_p3} as well as a small discontinuity between the two parts of the fitting functions Eq. \eqref{left_fit} and Eq. \eqref{right_fit} lead to a larger variation of $|M \psi^A|^2$ on the logarithmic scale of panel (a) of this figure. Despite these variations, which can be understood systematically, the stated property that $|M \psi^A|^2$ is significantly smaller than $|M \psi^B|^2$ holds true. (b) The parameter $\alpha$ in Eq. \eqref{bloch_states}, at which the low contrast takes place. The other spinor angle $\varphi$ is always determined to zero in our iterations. The lower figure is initial spin orientation parameter $\alpha$ of electron for the lowest contrast. The value of $\alpha$  at the left ($q_3=0$) and right ($q_3=1$) bounds match the analytic considerations in sections \ref{sec:consistency_considerations} and \ref{sec:low_electron_momentum}.}
 \label{M_S_p3}
\end{figure}

For demonstrating explicitly that spin-dependent diffraction takes place along the fitted line in Fig. \ref{1/theta_p3}, we display $|M \psi^A|^2$ and $|M \psi^B|^2$ as function of $q_3$ in Fig. \ref{M_S_p3}(a). The laser polarization $\vec A_{0,l}$ in Eq. \eqref{amplitude_r} at each value of $q_3$ in Fig. \ref{M_S_p3} is set according to the functional dependence $\theta(q_3)$ of the fitting functions in Eq. \eqref{eq:fit_functions}. The parameters $\alpha$ and $\varphi$ for the initial electron spin polarization in Fig. \ref{M_S_p3}(a) are determined as in the same way as for the $C'(M)$ minimum iteration in Figs. \ref{con_095_105} to \ref{1/theta_p3}, where we find that the spinor angle $\varphi(q_3)$ of the spinors \eqref{bloch_states} is always zero. The function of the other angle $\alpha(q_3)$ is displayed in Fig. \ref{M_S_p3}(b). We see in Fig. \ref{M_S_p3}(a) that $|M \psi^A|^2$ is several orders of magnitude smaller than $|M \psi^B|^2$. In other words, for each transverse momentum $q_3$, with $q_2=0$, we are able to find laser polarizations $\vec A_{0,r}$, $\vec A_{0,l}$ and an initial electron spin configuration, such the spin propagation matrix $M$ in the form of Eq. \eqref{Compton_formula} assumes matrix entries for which the two orthogonal polarizations $\psi^A$ and $\psi^B$ have a significantly different diffraction probability. This confirms the desired spin-dependent diffraction effect. We point out that Fig. \ref{M_S_p3} presents explicit parameters for the laser polarization and the electron spin polarization, such that the spin-dependent diffraction effect emerges.

\section{Discussion\label{sec:discussion}}

\subsection{Consistency considerations\label{sec:consistency_considerations}}

We refer back to the scenario of references \cite{ahrens_2020_two_photon_bragg_scattering,ahrens_2017_spin_non_conservation} at $q_3=1$, which is discussed in section \ref{sec:method_demonstration}, corresponding to the right end of the $x$-axis in Figs. \ref{1/theta_p3} and \ref{M_S_p3}. The fitting function \eqref{right_fit} at this right end evaluates to $1/\theta=4.005 /\pi$. This matches the value $\theta=\pi/4$, which was used in section \ref{sec:method_demonstration} and confirms the fitting procedure. Regarding the determined electron polarization, we mention that spin-dependent diffraction is discussed for the angle $\alpha=7 \pi/4 + 2\pi$ in references \cite{ahrens_2020_two_photon_bragg_scattering,ahrens_2017_spin_non_conservation}, where we skip the additional $2\pi$, as the contrast is $2 \pi$ periodic with respect to $\alpha$, as explained in appendix \ref{sec:iterative_algorithm}. The remaining value $7 \pi/4$ for the angle $\alpha$ matches the determined value in Fig. \ref{M_S_p3}(b), at $q_3=1$, and confirms our iterative algorithm.

\subsection{Limit for low transverse electron momenta\label{sec:low_electron_momentum}}

In analogy to the above section, we would like to do a similar analysis for the momentum $q_3=0$. This is of interest, as a low transverse electron momentum can allow for longer interaction times, for the interaction between laser and electron in the Kapitza-Dirac effect. Following the procedures in references \cite{ahrens_2020_two_photon_bragg_scattering,ahrens_2017_spin_non_conservation}, we perform a Taylor series expansion of the Compton tensor \eqref{Compton_tensor} around the point of transverse momenta $q_2=q_3=0$ up to second order in products of the variables $q_L$, $q_2$ and $q_3$. We obtain the matrix entries
\begin{align}
 M^{s's;22}&=\left(1-2q^2_2+\frac{1}{2}q_L^2\right)\mathds{1}-\frac{i}{2}q_{3}q_{L}\sigma_{2}-\frac{3i}{2}q_{2}q_{L}\sigma_{3}\nonumber\\
 M^{s's;23}&=-2q_{2}q_{3}\mathds{1}-iq_L\sigma_{1}+iq_{2}q_{L}\sigma_{2}-iq_{3}q_{L}\sigma_{3}\nonumber\\
 M^{s's;32}&=-2q_{2}q_{3}\mathds{1}+iq_L\sigma_{1}+iq_{2}q_{L}\sigma_{2}-iq_{3}q_{L}\sigma_{3}\nonumber\\
 M^{s's;33}&=\left(1-2q^2_3+\frac{1}{2}q_L^2\right)\mathds{1}+\frac{3i}{2}q_{3}q_{L}\sigma_{2}+\frac{i}{2}q_{2}q_{L}\sigma_{3}\,.\label{eq:taylor_expansion_compton_tensor}
\end{align}
On the basis of this expansion, the subsequent argument about the spin- and polarization dependent interaction is carried out in a calculation up to leading order in the small parameters $q_i$. For the polarization vectors in Eq. \eqref{amplitude_r} we choose the angle $\theta$ such that $\sin \theta$ evaluates to $q_L$. This is roughly the case for $\theta\approx q_L=0.02$. We obtain the polarizations $\vec A_{0,l}=\vec e_2 + i q_L \vec e_3$ and $\vec A_{0,r}=\vec e_3$ for the left and right propagating laser beams. Together with the Taylor expansion \eqref{eq:taylor_expansion_compton_tensor}, these polarizations evaluate in the Compton scattering formula \eqref{Compton_formula} as
\begin{equation}
M \approx - i q_L \mathds{1} - i q_L \sigma_1 
 = - i q_L
\begin{pmatrix}
 1 & 1 \\ 1 & 1
\end{pmatrix}\,.\label{eq:zero_transverse_momentum_spin_propagation}
\end{equation}
We see that the eigenvectors of $\sigma_1$
\begin{equation}
 \psi^A=\frac{1}{\sqrt{2}}
\begin{pmatrix}
 -1 \\ 1
\end{pmatrix}\,,\qquad
 \psi^B=\frac{1}{\sqrt{2}}
\begin{pmatrix}
 -1 \\ -1
\end{pmatrix}\label{eq:x_polarization_spinors}
\end{equation}
produce a zero contrast $\mathcal{C}(M)=0$. The states \eqref{eq:x_polarization_spinors} correspond to the spinors \eqref{bloch_states} with parameters $\alpha=3 \pi/2$ and $\varphi=0$. This matches the algorithmically determined value for $\alpha$ and $\varphi$ at the left end of the $x$-axis in Fig. \ref{M_S_p3}(b), for the chosen photon energy $q_L=0.02$. Similarly, the left side of Fig. \ref{1/theta_p3} is confirmed, since we have chosen $1/\theta\approx1/q_L=50$. The fitting function \eqref{left_fit} is consistently evaluating to the value $1/\theta(0)=50.13$\,.

\begin{figure}
 \includegraphics[width=0.5\textwidth]{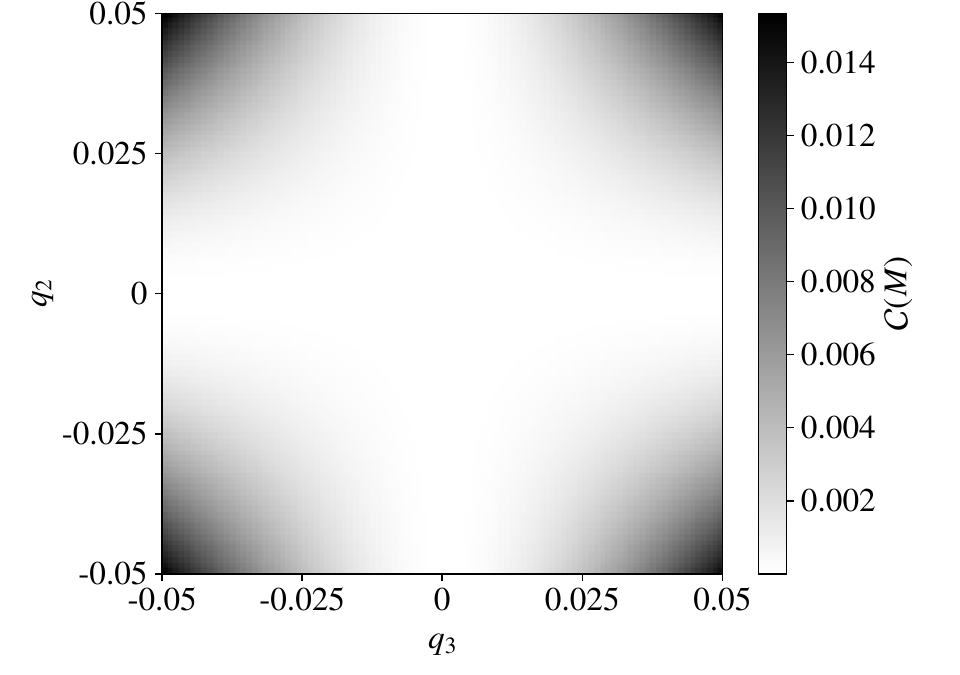}
 \caption{Contrast around the momentum $q_2=q_3=0$. We display the contrast similarly as in Fig. \ref{con_095_105}, but now for the range $q_3\in[-0.05,0.05]$, such that it is centered around the momentum $q_3=0$. The ellipticity parameter of the left propagating laser beam is set to the value the value $\theta=1/50$. We find a contrast close to zero at the momenta $q_2=q_3=0$.}
 \label{con_005_005}
\end{figure}

\begin{figure}
 \includegraphics[width=0.5\textwidth]{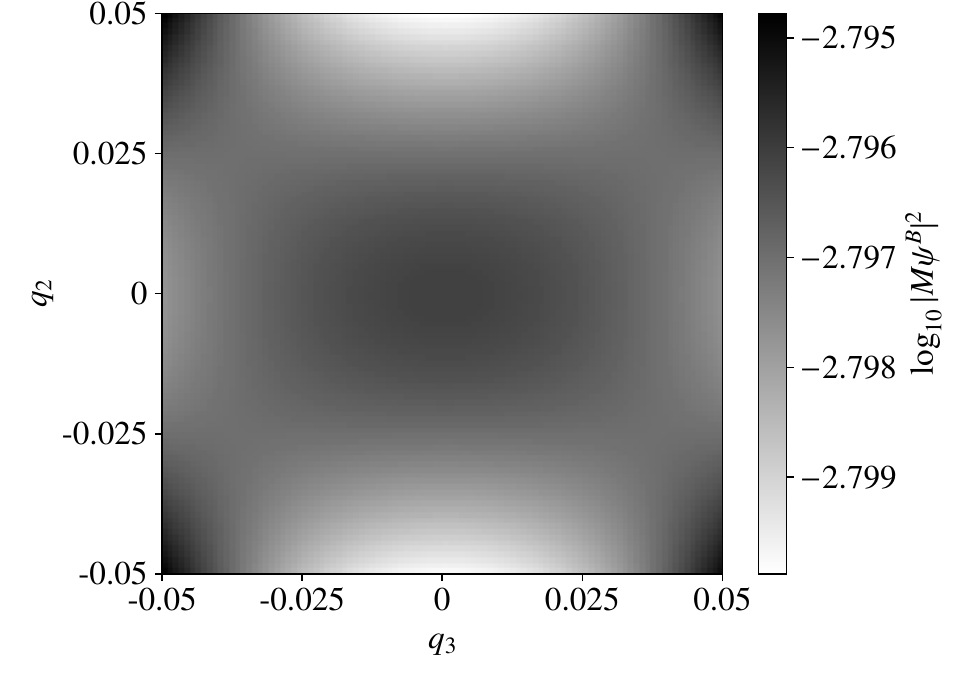}
 \caption{\label{M_005_005}Diffraction probability around the momentum $q_2=q_3=0$. Similarly to the contrast over the range $q_3\in[-0.05,0.05]$ in Fig. \ref{con_005_005} as a modification of Fig. \ref{con_095_105}, we display here the diffraction probability $|M \psi_B|^2$ over the range $q_3\in[-0.05,0.05]$ and \ $\theta=1/50$, as a modification of Fig. \ref{M_095_105}. We see again a non-vanishing diffraction probability, which, in combination with the low contrast in Fig. \ref{con_005_005}, implies the existence of spin-dependent diffraction at $q_2=q_3=0$.}
\end{figure}

We further confirm our considerations by displaying the contrast and diffraction probability as a function of the momenta $q_2$ and $q_3$ around the point $q_2=q_3=0$ in Figs. \ref{con_005_005} and \ref{M_005_005}, respectively. The procedure is similar to the the variation of momenta in Figs. \ref{con_095_105} and \ref{M_095_105}, with the difference, that the momentum interval at the $x$-axis of Figs. \ref{con_005_005} and \ref{M_005_005} is now $q_3\in [-0.05,0.05]$, and the ellipticity parameter of the left propagating laser beam is set to $1/\theta=50$, as discussed above. We can see in Fig. \ref{con_005_005} a contrast close to zero, where Fig. \ref{M_005_005} implies a non-vanishing diffraction probability for the discussed spin dynamics. This confirms our conclusion from above: Spin-dependent diffraction is possible for the transverse momenta $q_2=q_3=0$, with laser and electron polarization parameters $1/\theta=50$ and $\alpha=3 \pi/2$, $\varphi=0$ as implied on the left ($q_3=0$) of Figs. \ref{1/theta_p3} and \ref{M_S_p3}, for the case of the laser frequency $q_L=0.02$. Note, that a code example for the iterative algorithm with the mentioned parameters is provided in the supplemental material, for the specific case of the momenta $q_2=q_3=0$. We also point out, that we give a thorough description about the spin configuration of the electron spin in Fig. \ref{con_005_005} and \ref{M_005_005}, similarly as done for Figs. \ref{con_095_105} and \ref{M_095_105} in Appendix \ref{sec:spin_expectation_value}. 

\section{Conclusion and Outlook\label{sec:outlook}}

In this study, we are introducing the contrast on the basis of Eq. \eqref{contrast}. For a given spin propagation matrix, the initial electron spin polarization is to be optimized such that contrast reaches its minimum. This definition can serve as a measure for quantifying the spin-dependent diffraction in the Kapitza-Dirac effect. With the help of an iterative algorithm, as described in appendix \ref{sec:iterative_algorithm}, which optimizes any given spin propagation matrix $M \in \mathbb{C}^{2\times 2}$ regarding the spin-dependent diffraction effect, we are able to systematically search for the optimal parameters at which a spin-dependent Kapitza-Dirac effect can take place. Specifically, we are demonstrating in Figs. \ref{1/theta_p3} and \ref{M_S_p3}, that spin-dependent diffraction with a contrast close to zero ($\mathcal{C}(M)\approx 0$) can always be achieved for any transverse momentum between 0 and $1mc$ along the direction of the linearly polarized laser beam. The zero contrast situation takes place when one of the counter-propagating beams in the Kapitza-Dirac effect is polarized along the transverse electron momentum, whereas the polarization of the other beam is elliptically polarized according to the polarization \eqref{amplitude_r} with ellipticity  $\theta$ given by Eq. \eqref{eq:fit_functions} and related fit parameters \eqref{eq:fit_parameters}. Therefore, our results demonstrate that spin-dependent two-photon Kapitza-Dirac scattering in the Bragg regime is possible for arbitrary small transverse electron momenta, implying longer possible interaction times of the laser with the electron. Since the diffraction probability is growing quadratically with the effective interaction time between the laser and the electron, our finding can enhance the count-rate in an experimental setup or, analogously, lower the necessary laser intensity which is required in such an experiment.

In the future, it would be interesting to apply the discussed spin optimization procedure to the spin propagation of the Kapitza-Dirac effect with strongly focused laser beams \cite{ahrens_guan_2022_beam_focus_longitudinal}. Another application of the contrast technology is an investigation of the dimenionality of parameter space (ie. numbers of parameters which can be varied independently from each other), such that spin-dependent electron diffraction can still be observed.

\begin{acknowledgments}
The work was supported by the National Natural Science Foundation of China (Grant No. 11975155).
\end{acknowledgments}

\appendix

\section{The contrast iteration algorithm\label{sec:iterative_algorithm}}
For the determination of the contrast $\mathcal{C}(M)$ we are interested in the minimization of the function $\mathcal{C}'(M)$, defined in Eq. \eqref{contrast}, as a function of $\alpha$ and $\varphi$, for a given complex $2\times 2$ matrix $M$. For the minimization we make use of the Newton's method in two dimensions \cite{Nocedal_Wright_2006_numerical_optimization}, with the iteration step
\begin{equation}
 \left(
 \begin{array}{c}
  \alpha_{n+1} \\
  \varphi_{n+1}
 \end{array}
 \right) =
 \left(
 \begin{array}{c}
  \alpha_n \\
  \varphi_n
 \end{array}
 \right)
 - H(\alpha_n,\varphi_n)^{-1} \vec g(\alpha_n,\varphi_n)\,,\label{eq:iteration_ansatz}
\end{equation}
where $H(\alpha_n,\varphi_n)^{-1}$ is the inverse of the Hessian matrix
\begin{equation}
H(\alpha_n,\varphi_n)=
\left.
\begin{bmatrix}
 \partial_\alpha^2 \mathcal{C}'(M) & \partial_\alpha \partial_\varphi \mathcal{C}'(M) \\
 \partial_\varphi \partial_\alpha \mathcal{C}'(M) & \partial_\varphi^2 \mathcal{C}'(M)
\end{bmatrix}
\right|_{\alpha = \alpha_n \atop \varphi = \varphi_n }
\end{equation}
and $\vec g(\alpha_n,\varphi_n)$ is the gradient
\begin{equation}
\vec g(\alpha_n,\varphi_n)=
\left.
\begin{bmatrix}
 \partial_\alpha \mathcal{C}'(M) \\
 \partial_\varphi \mathcal{C}'(M)
\end{bmatrix}
\right|_{\alpha = \alpha_n \atop \varphi = \varphi_n }\,.
\end{equation}
For avoiding numerical inaccuracy, we have manually computed and implemented the derivatives $\partial_\alpha \mathcal{C}'(M)$, $\partial_\varphi \mathcal{C}'(M)$, $\partial_\alpha^2 \mathcal{C}'(M)$, $\partial_\varphi^2 \mathcal{C}'(M)$ and $\partial_\alpha \partial_\varphi \mathcal{C}'(M)$. The calculation of the derivatives and also an executable example of our algorithm is provided in the supplemental material.

Note, that the spinors \eqref{bloch_states} are periodic, with periodicity interval $\alpha \in [0,4 \pi]$, $\varphi \in [0,2 \pi]$. Further, the spinors receive a minus sign over the range $\alpha \in [2\pi,4 \pi]$, which drops out in the absolute value of the contrast definition \eqref{contrast}. We also receive the same spinors on the substitution $\varphi \rightarrow \varphi + \pi$ and $\alpha \rightarrow 2 \pi - \alpha$. Therefore, for avoiding ambiguities, we restrict the parameters of the spinors \eqref{bloch_states} to the bounds
\begin{equation}
 \alpha \in [0,2 \pi]\,,\qquad \varphi \in [0, \pi]\,. \label{eq:parameter_bounds}
\end{equation}

The computation of the contrast is implemented in the following way:
\begin{itemize}
 \item For given polarizations $\vec A_{0,l}$ and $\vec A_{0,r}$ of the counterpropagating laser beams we compute the spin propagation matrix $M$ of the Kapitza-Dirac effect from the Compton scattering formula \eqref{Compton_formula}.
 \item For avoiding the iteration into a wrong local minimum, we first iterate over all value pairs $(\alpha,\varphi)$ on a grid of equal spacing, with 126 different grid points for $\alpha$ and 63 grid points for $\varphi$. We assume the pair $(\alpha,\varphi)$ with the lowest value in Eq. \eqref{contrast} to be near the global minimum and use it as initial values $\alpha_0$ and $\varphi_0$ for the iteration.
 \item With the initial values set, we iterate Newton's method in two dimensions \eqref{eq:iteration_ansatz}, where the Hessian matrix $H(\alpha_n,\varphi_n)$ in and the gradient $\vec g(\alpha_n,\varphi_n)$ are computed for each iteration step.
 \item After each iteration step, we check whether $\alpha_n$ and $\varphi_n$ are inside of the restricted bounds \eqref{eq:parameter_bounds} and use the above discussed periodicity properties, to reassign the parameters back to the bounds \eqref{eq:parameter_bounds}, in case they are not within these bounds.
 \item We stop the iteration, when the gradient is low $|g(\alpha_n,\varphi_n)|<10^{-15}$, the Hessian matrix is nearly singular (ie. inversion is getting more inaccurate) $\det[H(\alpha_n,\varphi_n)]<10^{-20}$, or the convergence doesn't take place after 80 iterations.
\end{itemize}

\section{The spin expectation values of the electron in spin-dependent diffraction\label{sec:spin_expectation_value}}

We would like to provide information about the electron spin configuration before and after Kapitza-Dirac scattering for the situation in Figs. \ref{con_095_105} and Figs. \ref{M_095_105} and likewise for the situation in Figs. \ref{con_005_005} and Fig. \ref{M_005_005}. We do this by evaluating the spin expectation value
\begin{subequations}\label{eq:spin_expectation_values}%
\begin{equation}%
 \frac{\braket{\psi^B|S_i|\psi^B}}{\braket{\psi^B|\psi^B}}\label{eq:spin_expectation_value_before_scattering}
\end{equation}%
of the spin 1/2 spin operator $S_i=\hbar \sigma/2$ for the contrast optimized values $\alpha$ and $\varphi$ over the parameter range of transverse electron momenta of the figures. The spin expectation value in Eq. \eqref{eq:spin_expectation_value_before_scattering} provides the spin polarization direction of the electron before scattering, whereas the spin expectation value of the electron after scattering is given by%
\begin{equation}%
 \frac{\braket{M \psi^B|S_i|M \psi^B}}{\braket{M \psi^B|M \psi^B}}\,.\label{eq:spin_expectation_value_after_scattering}
\end{equation}%
Normalizing the spin expectation value by the norm of the spin state in Eqs. \eqref{eq:spin_expectation_value_before_scattering} and \eqref{eq:spin_expectation_value_after_scattering} ensures the length of the displayed spin vector to be $\hbar/2$. The spin polarization direction of $\psi^A$ before scattering is pointing in the opposite direction of $\psi^B$ in Eq. \eqref{eq:spin_expectation_value_before_scattering}. For the situation of the spin state $M \psi^A$ after scattering, we consider the quantity%
\begin{equation}%
 \frac{\braket{M \psi^A|S_i|M \psi^A}}{\braket{M \psi^B|M \psi^B}}\,,\label{eq:spin_hindered_expectation_value_after_scattering}
\end{equation}%
\end{subequations}%
normalized by $\braket{M \psi^B|M \psi^B}$, since the denominator $\braket{M \psi^A|M \psi^A}$ would lead to the situation of a numerically unstable, lifted singularity, in case of a zero contrast.

\begin{figure}
  \includegraphics[width=0.5\textwidth]{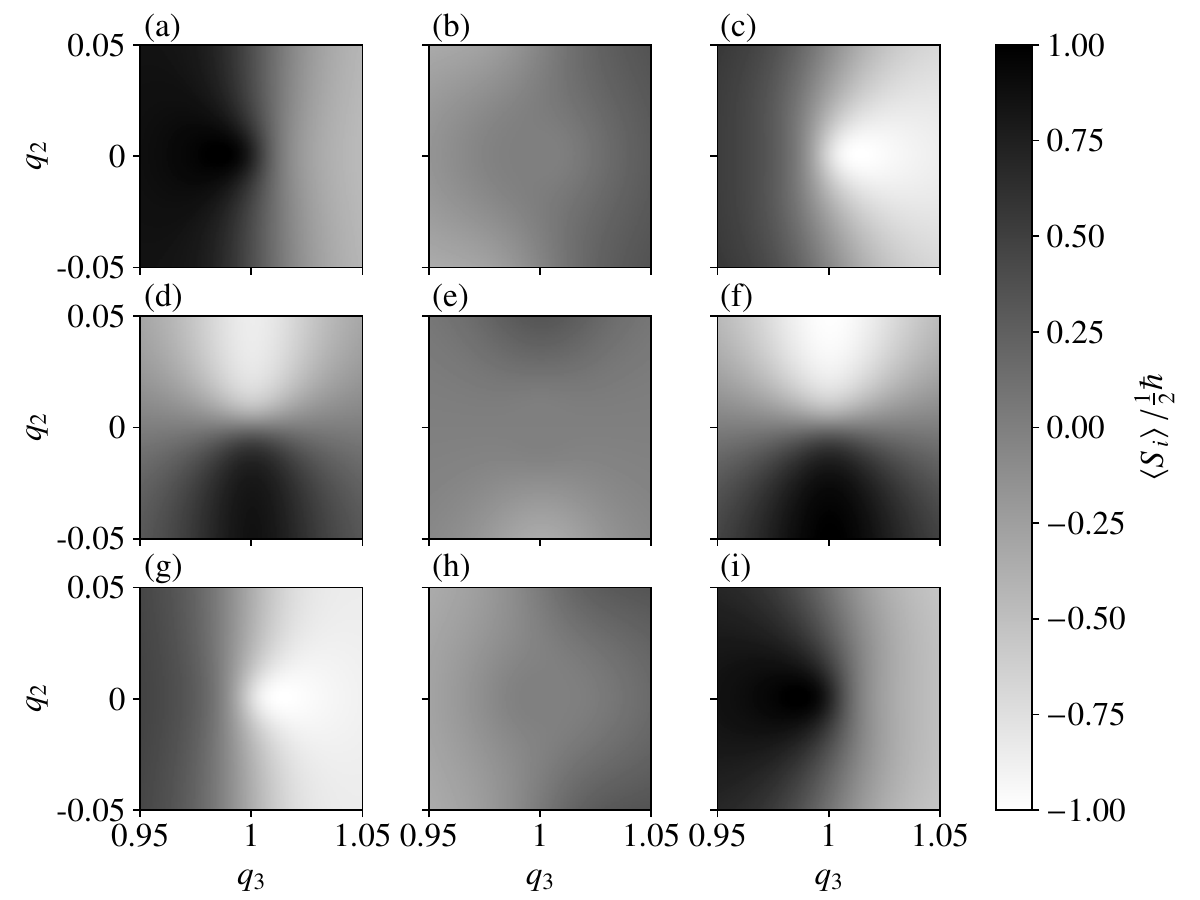}
  \caption{Display of spin expectation value \eqref{eq:spin_expectation_values} for the setup in Figs. \ref{con_095_105} and \ref{M_095_105}. Fig. \ref{spin_expectation_fig12} (a), (d), (g) show the spin expectation value for the spin state $\psi^B$, as in Eq. \eqref{eq:spin_expectation_value_before_scattering}. Fig. \ref{spin_expectation_fig12} (b), (e), (h) are the normalized spin expectation values \eqref{eq:spin_hindered_expectation_value_after_scattering} for the spin state $M\psi^A$, with the normalization factor $(\braket{M\psi^B|M\psi^B})^{-1}$. Fig. \ref{spin_expectation_fig12} (c), (f), (i) are the normalized spin expectation value for the spin state $M\psi^B$, with the normalization factor $(\braket{M\psi^B|M\psi^B})^{-1}$, as in Eq. \eqref{eq:spin_expectation_value_after_scattering}. Regarding the horizontal panel arrangement, the panels (a), (b), (c) are the $x$ components $\braket{S_1}$; (d), (e), (f) are the $y$ components $\braket{S_2}$; and (g), (h), (i) are the $z$ components $\braket{S_3}$. In the central region we see a spin-flip along the $\vec e_1 - \vec e_3$ direction. The $y$-component of the spin orientation almost unchanged, when diffraction occurs.
  \label{spin_expectation_fig12}}
\end{figure}

In Fig. \ref{spin_expectation_fig12} we display the spin-expectation values \eqref{eq:spin_expectation_values} for the situation in Figs. \ref{con_095_105} and Figs. \ref{M_095_105}, where the panels (a), (d), (g) are the three components of Eq. \eqref{eq:spin_expectation_value_before_scattering}, (b), (e), (h) are the three components of Eq. \eqref{eq:spin_hindered_expectation_value_after_scattering} and (c), (f), (i) are the three components of Eq. \eqref{eq:spin_expectation_value_after_scattering}. We note, that the $y$-component of the electron's spin polarization shows no significant change before [Fig. \ref{spin_expectation_fig12}(d)] and after [Fig. \ref{spin_expectation_fig12}(f)] scattering. Also the $x$ and $z$ components in Figs. \ref{spin_expectation_fig12}(a) and \ref{spin_expectation_fig12}(g) before diffraction appear very similar to Figs. \ref{spin_expectation_fig12}(c) and \ref{spin_expectation_fig12}(i) after diffraction. Only in the central region, we observe a spin-flip along the $\vec e_1 - \vec e_3$ direction, which is consistent with the description in references \cite{ahrens_2017_spin_non_conservation,ahrens_2020_two_photon_bragg_scattering} and our considerations in section \ref{sec:consistency_considerations}.

\begin{figure}
  \includegraphics[width=0.5\textwidth]{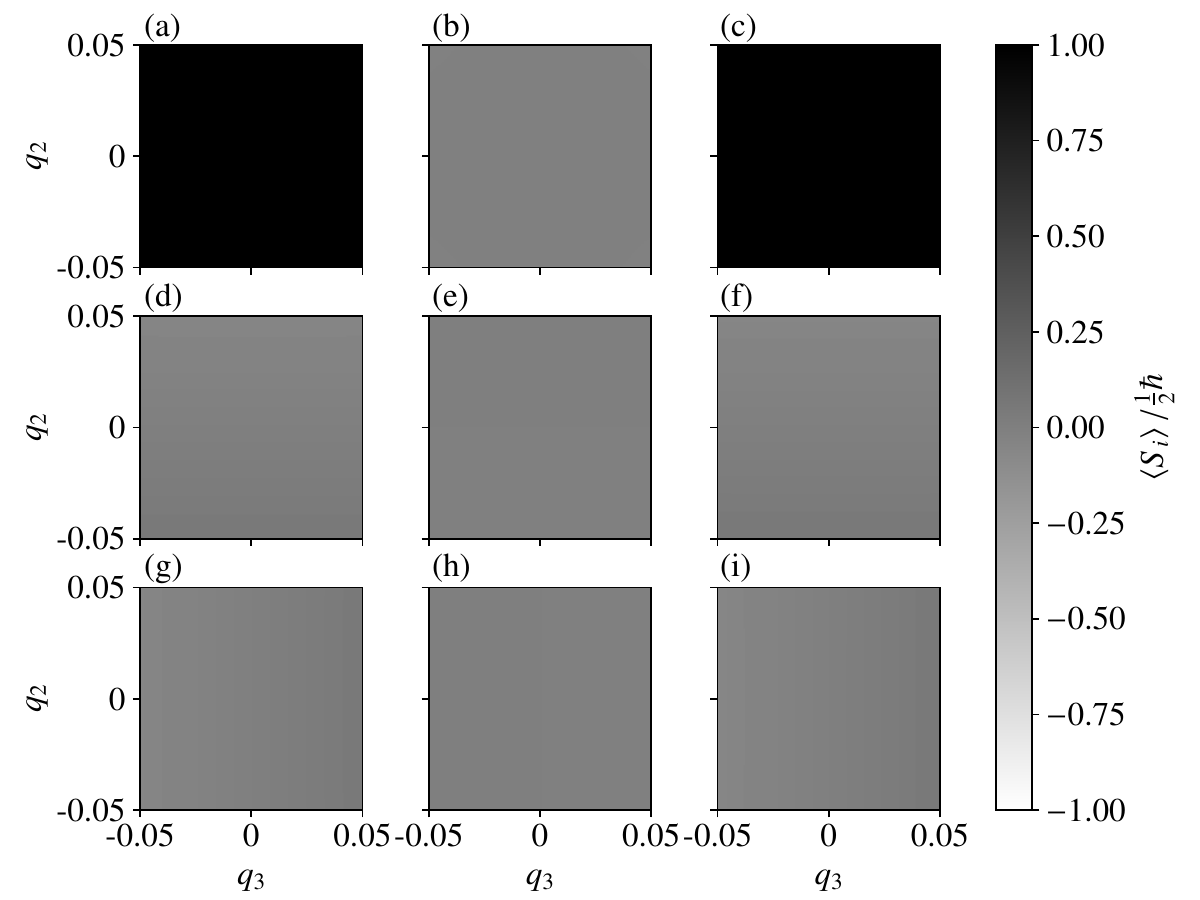}
  \caption{Display of spin expectation value \eqref{eq:spin_expectation_values} for the setup in Figs. \ref{con_005_005} and \ref{M_005_005}. The panel arrangement is as in Fig. \ref{spin_expectation_fig12}. In the case of diffraction a $+x$ polarized electron is scattered into a $+x$ polarized state.\label{spin_expectation_fig67}}
\end{figure}

We also display the spin-expectation values 
\eqref{eq:spin_expectation_values} for the situation in Figs. \ref{con_005_005} and Fig. \ref{M_005_005} in Fig.  
\ref{spin_expectation_fig67}, with the same panel arrangement as in Fig. \ref{spin_expectation_fig12}. We clearly observe a spin polarization of $\psi^B$ and $M \psi^B$ along the $x$ direction, corresponding to our considerations in section \ref{sec:low_electron_momentum}, with $\psi^B$ in Eq. \eqref{eq:x_polarization_spinors} being the $\sigma_1$ eigenstate with eigenvalue +1 and after applying to the spin-propagation matrix \eqref{eq:zero_transverse_momentum_spin_propagation} remaining a $\sigma_1$ eigenstate with eigenvalue +1. In both, Fig. \ref{spin_expectation_fig12} and \ref{spin_expectation_fig67}, the quantity \eqref{eq:spin_hindered_expectation_value_after_scattering}, in panels (b), (e), (h) is very close to zero, consistent with the contrast in Fig. \ref{con_095_105} and in particular in Fig. \ref{con_005_005}.

\bibliography{bibliography}

\end{document}